\documentclass[%
reprint,
 amsmath,amssymb,
 aps,
prb,
floatfix
]{revtex4-2}

\usepackage{graphicx}
\usepackage{dcolumn}
\usepackage{bm}
\usepackage{hyperref}
\usepackage{xcolor}
\usepackage{physics}

\usepackage{enumitem}
\usepackage{amssymb}

\def\mb#1{\mathbf{#1}}

\begin{document}


\title{Examining the microscopic origin of a computationally inexpensive thermal-conductivity-based indicator for phonon hydrodynamics}

\author{Nikhil Malviya}
\author{Navaneetha K. Ravichandran}
\email{navaneeth@iisc.ac.in}
\affiliation{%
 Department of Mechanical Engineering, Indian Institute of Science, Bangalore 560012, India
 }%

\date{\today}

\begin{abstract}
Hydrodynamic heat flow, where out-of-equilibrium phonons collectively drift in response to a temperature gradient, has attracted renewed interest following its experimental observation in graphite from cryogenic to room temperatures. To rapidly screen for other materials exhibiting this unconventional non-Fourier regime, the computationally inexpensive thermal conductivity ratio $\kappa_{\text{LPBE}}$/$\kappa_{\text{RTA}}$ obtained from a complete solution of the linearized Peierls–Boltzmann equation (LPBE) for phonon transport and the relaxation time approximation (RTA) for phonon decay, has often been used as an indicator for phonon hydrodynamics. Yet, a clear connection between $\kappa_{LPBE}/\kappa_{RTA}$ and the signatures of drifting hydrodynamic phonon populations has not been established in the literature. Here we show that $\kappa_{LPBE}/\kappa_{RTA}$ directly correlates with the microscopic hydrodynamic signatures arising from the spectral properties of the phonon collision operator, which is often computationally expensive to compute, thus establishing the former as a reliable low-cost indicator for phonon hydrodynamics. On the other hand, other indicators in the literature that are derived only from the phonon scattering rates do not correlate with $\kappa_{LPBE}/\kappa_{RTA}$, and so, are inadequate to predict phonon hydrodynamics. Our study also reveals that $\kappa_{\text{LPBE}}/\kappa_{\text{RTA}}$, and therefore the strength of hydrodynamic signatures, decrease with increasing Brillouin zone (BZ) sampling density for several ultrahigh-$\kappa$ materials at low temperatures, thus underscoring the need for careful BZ sampling for robust predictions of phonon hydrodynamics. Our work justifies the use of $\kappa_{\text{LPBE}}/\kappa_{\text{RTA}}$ as a computationally inexpensive indicator of phonon hydrodynamics, thus enabling accelerated search for new materials that exhibit such unconventional heat flow regimes beyond classical Fourier's law.
\end{abstract}
\maketitle

\clearpage
Heat in semiconductors is predominantly carried by the quantized eigenmodes of the harmonic crystal potential called phonons. These phonons scatter among themselves due to the crystal anharmonicity, resulting in a resistance to heat flow and a finite thermal conductivity ($\kappa$). These scattering processes can be classified as momentum-conserving Normal scattering (N-scattering) and momentum-dissipative Umklapp scattering (U-scattering) processes. In materials where N-scattering is much stronger than U-scattering, heat can flow like a viscous fluid rather than a diffusive gas, resulting in an unconventional heat flow regime called the hydrodynamic heat flow~\cite{guyer_thermal_1966, hardy_phonon_1970, lee_hydrodynamic_2015, cepellotti_phonon_2015, simoncelli_generalization_2020}, which cannot be described by the commonly-used Fourier's heat diffusion law. This hydrodynamic heat flow phenomenon has captured the attention of the research community owing to its ability to introduce a directionality to heat flow through thermal rectification --- a thermal Tesla valve~\cite{huang_graphite_2024} and its potential applications in thermal cloaking and shielding~\cite{shi_nonresistive_2019, chen_non-fourier_2021}.

The transient manifestation of this hydrodynamic regime called the second sound, where heat propagates at speeds comparable to that of sound, has been observed in sodium fluoride~\cite{jackson_second_1970, jackson_thermal_1971} and bismuth~\cite{narayanamurti_observation_1972} at temperatures below 20~K, and recently in sapphire~\cite{kawabata_phonon_2025}, germanium~\cite{beardo_observation_2021} and graphite~\cite{huberman_observation_2019, ding_observation_2022, xie_room-temperature_2026} from 100~K to 300~K. These experimental realizations of hydrodynamic heat flow open up new possibilities to control and manipulate heat in several classical and quantum electronic applications. To rapidly discover new materials that exhibit this unconventional heat flow regime, the ratio of $\kappa$ obtained from the complete solution of the governing equation for phonon transport --- the linearized Peierls-Boltzmann equation (LPBE) to that obtained from the relaxation time approximation (RTA) to the LPBE, is often used as a computationally inexpensive indicator for phonon hydrodynamics~\cite{sjakste_occurrence_2024, zhang_hydrodynamic_2020, lindsay_perspective_2019, duan_hydrodynamically_2022, markov_hydrodynamic_2018, torres_thermal_2019}. While it is generally understood that the ratio $\kappa_{LPBE}/\kappa_{RTA}$ is correlated with the relative strengths of N- and U-scattering processes --- a requirement for phonon hydrodynamics, a clear connection between this ratio and the microscopic signatures of collectively drifting hydrodynamic phonon populations has not been established in the literature.

Here, we demonstrate a direct correlation between the ratio $\kappa_{LPBE}/\kappa_{RTA}$ and the hydrodynamic signatures arising from the spectral properties of the phonon collision operator, thus establishing the former as a reliable indicator for phonon hydrodynamics. We show that this strong correlation originates from the off-diagonal terms of the phonon collision matrix that is dominated by N-scattering events. Therefore, other indicators relying only on the diagonal part, such as the ratio of average N- and U-scattering rates~\cite{zhang_hydrodynamic_2020, duan_hydrodynamically_2022, markov_hydrodynamic_2018}, cannot capture this correlation. Since the eigenspectrum of the phonon collision operator is often expensive to compute, our work presents a microscopically-founded justification for using $\kappa_{LPBE}/\kappa_{RTA}$ as a low-cost indicator for phonon hydrodynamics, that will accelerate data-driven computational searches for material alternatives and experimental conditions to realize this unconventional heat flow regime.

To obtain $\kappa$, we solve the steady state LPBE for the linearized non-equilibrium deviational phonon distribution function $\left[f_{\lambda} = \frac{\tilde{f}-f_{\lambda}^{0}}{\sqrt{f_{\lambda}^{0}\left(f_{\lambda}^{0}+1\right)}}\right]$ given by:
\begin{align}
    \frac{\mathbf{v}_{\lambda} \cdot \nabla f^{0}_{\lambda}}{\sqrt{f^0_\lambda\left(f^0_\lambda + 1\right)}} = & - \sum_{\lambda_{1}} \Omega_{\lambda \lambda_{1}} f_{\lambda_{1}}
    \label{eq:steady_state_lpbe}
\end{align}
where $\tilde{f}_{\lambda}$ is the total non-equilibrium phonon distribution function, $\mathbf{v}_{\lambda}$ is the group velocity of the phonon mode $\lambda \equiv \left[ \mathbf{q}, j \right]$ with wave vector $\mathbf{q}$, polarization $j$ and frequency $\omega_\lambda$, $f_{\lambda}^{0} = \frac{1}{\exp\left[\frac{\hbar\omega_\lambda}{k_BT_0}\right]-1}$ is the corresponding equilibrium Bose-Einstein distribution at a temperature $T_0$ and $\mb{\Omega}$ is the collision matrix. We consider three-phonon, four-phonon and phonon-isotope scattering processes while constructing $\mb{\Omega}$ from first principles, as described in Refs.~\cite{ravichandran_unified_2018, ravichandran_phonon-phonon_2020, ravichandran_elasticity_2026} and Supplementary Note 1. Once $\mb{\Omega}$ is computed, Eq.~\ref{eq:steady_state_lpbe} is solved for $f_\lambda$ using iterative solvers~\cite{broido_intrinsic_2007, fugallo_ab_2013}, which have become computationally inexpensive in recent years due to the improvements in computing hardware and algorithms. From $f_{\lambda}$, we obtain the heat flux $\mathbf{J} = \frac{1}{V} \sum_{\lambda} \hbar \omega_{\lambda} \mathbf{v}_{\lambda} \sqrt{f^{0}_{\lambda} \left( f^{0}_{\lambda} + 1 \right)} f_{\lambda}$ for a crystal volume $V$, and cast it into a linear response relation: $\mathbf{J} = - \kappa_{LPBE} \nabla T$ to obtain $\kappa_{LPBE}$.

In general, the collision matrix, $\mb{\Omega}$, is dense. However, if the off-diagonal terms of $\mb{\Omega}$ are sufficiently small, then $\Omega_{\lambda \lambda_{1}} \approx \Delta_{\lambda \lambda_{1}} / \tau_{\lambda}$, where $\Delta_{\lambda \lambda_{1}}$ is the Kronecker delta and $\tau_{\lambda}$ is the phonon relaxation time. With this simplification, called the RTA (introduced earlier), Eq.~\ref{eq:steady_state_lpbe} can be solved in closed-form to obtain $\kappa_{RTA, \alpha} = \sum_{\lambda} C_{0 \lambda} v_{\alpha \lambda}^{2} \tau_{\lambda}$ with $C_{0\lambda}$ being the volumetric phonon heat capacity and $\alpha$ is the Cartesian index which is dropped moving forward due to the crystal symmetries of the materials considered here. Although the RTA is computationally efficient due to the dependence on the diagonal terms of $\mb{\Omega}$ only, it works reasonably well only for low-$\kappa$ semiconductors like silicon and germanium~\cite{malviya_failure_2023}, but underpredicts the $\kappa$ of ultrahigh-$\kappa$ materials like diamond, where the large off-diagonal terms of $\mb{\Omega}$ that induce strong coupling among the $f_\lambda$'s of different phonons in Eq.~\ref{eq:steady_state_lpbe} become important. In fact, these off-diagonal terms are dominated by N-scattering events as shown in Ref.~\cite{malviya_failure_2023}, which drives a collective drifting motion of out-of-equilibrium phonons, culminating in hydrodynamic heat flow.

To overcome the limitations of the RTA while retaining its computational efficiency, the Callaway approximation to the LPBE~\cite{callaway_model_1959, allen_improved_2013} has been explored as an alternative to predict $\kappa$ and phonon hydrodynamics in ultrahigh-$\kappa$ materials~\cite{guo_heat_2017, zhou_nonmonotonic_2017, ezzahri_thermal_2022, qian_analytical_2025}. In this approximation, the LPBE (Eq.~\ref{eq:steady_state_lpbe}) is simplified as:
\begin{align}
    \mathbf{v}_{\lambda} \cdot \nabla f^{0}_{\lambda} = & - \sum_{\lambda_{1}} \Omega^{\left(N\right)}_{\lambda \lambda_{1}} \left(\tilde{f}_{\lambda_{1}} - f^{*}_{\lambda_{1}}\right)\Delta_{\lambda\lambda_{1}} \nonumber \\ & - \sum_{\lambda_{1}} \Omega^{\left(U\right)}_{\lambda \lambda_{1}} \left(\tilde{f}_{\lambda_{1}} - f^{0}_{\lambda_{1}}\right)\Delta_{\lambda\lambda_{1}}
    \label{eq:steady_state_Callaway}
\end{align}
where $\mb{\Omega}^{\left(N\right)}$ and $\mb{\Omega}^{\left(U\right)}$ are the collision matrices for the N- and the U-processes respectively, $f^{*}_{\lambda} = \frac{1}{\exp\left[\frac{\hbar\omega_\lambda - \mb{q}\cdot\Lambda}{k_BT_0}\right]-1}$ is a drifting equilibrium distribution function with a drift velocity $\Lambda$ that is determined using a closure condition describing the total momentum conservation of a phonon system undergoing N-processes only, and $\Delta_{\lambda\lambda_1}$ ensures that only the diagonal terms of $\mb{\Omega}^{\left(N/U\right)}$ appear in Eq.~\ref{eq:steady_state_Callaway}, thus retaining the computational simplicity of the RTA. Although the Callaway approximation works reasonably well for several ultrahigh-$\kappa$ materials like diamond, boron nitride (BN) and boron phosphide (BP), it fails to predict the ultrahigh-$\kappa$ of two special materials --- boron arsenide (BAs) and boron antimonide (BSb)~\cite{malviya_failure_2023}, and fails to capture the hydrodynamic second sound signatures at low temperatures~\cite{malviya_callaway_2025}. Interestingly, both of these failures originate from the very feature that activates the hydrodynamic heat flow --- weak U-scattering processes due to the activation of the phonon scattering selection rules in the former~\cite{malviya_failure_2023} and low temperatures in the latter~\cite{malviya_callaway_2025}.

More importantly, $\kappa_{LPBE}$, $\kappa_{RTA}$ and $\kappa_{Callaway}$ do not provide any fundamental insight to predict the materials and conditions for phonon hydrodynamics, especially when U-processes are not negligible. Such insights are obtained directly from the spectral contributions of the eigenmodes of $\mb{\Omega}$ to the overall $\kappa$~\cite{hardy_phonon_1970, cepellotti_thermal_2016, malviya_efficient_2025}. These eigenmodes, also called relaxons~\cite{cepellotti_thermal_2016, simoncelli_generalization_2020}, form a complete orthonormal basis owing to the symmetric form of the collision operator $\Omega_{\lambda\lambda_1} = \Omega_{\lambda_1\lambda}$ and so, can be used to represent the solution of Eq.~\ref{eq:steady_state_lpbe} as $f_\lambda = \sum_i\vartheta_i\mathfrak{e}^{\left(i\right)}_\lambda$. With this eigenmode representation of $f_\lambda$, Eq.~\ref{eq:steady_state_lpbe} can be solved for the unknown coefficients $\{\vartheta_i\}$, and $\kappa_{LPBE}$ is obtained as $\kappa_{LPBE, \alpha} \equiv \kappa_{LPBE} = C_0\sum_{i}\left(\mathcal{V}^{0i}_\alpha\right)^{2}/\sigma_i$, where $\sigma_i$ is the eigenvalue and $\mathcal{V}^{0i}_\alpha = \sum_\lambda \mathfrak{e}^{\left(0\right)}_\lambda v_{\lambda, \alpha}\mathfrak{e}^{\left(i\right)}_\lambda$ is the velocity in the Cartesian direction $\alpha$ of the eigenmode $\mathfrak{e}^{\left(i\right)}$ of $\mb{\Omega}$, $C_0$ is the total volumetric heat capacity of the phonon system and $\mathfrak{e}^{\left(0\right)}$ represents the eigenmode of $\mb{\Omega}$ corresponding to the equilibrium distribution~\cite{cepellotti_phonon_2015, hua_space-time_2020, malviya_efficient_2025}. While there are as many eigenmodes of $\mb{\Omega}$ as there are phonon modes in the discretized Brillouin zone (BZ), in the hydrodynamic regime, the entire contribution to $\kappa_{LPBE}$ originates only from a few of them with nearly-degenerate eigenvalues, representing a strong collective drifting motion~\cite{hardy_phonon_1970}. This requirement is also consistent with the limiting case of negligible U-scattering ($\mb{\Omega} \approx \mb{\Omega}^{\left(N\right)}$), where three drifting eigenmodes of $\mb{\Omega}^{\left(N\right)}$ with vanishing eigenvalues contribute entirely to $\kappa$, resulting in a purely hydrodynamic heat flow in three-dimensional (3D) crystals~\cite{pitaevskii_physical_2012}.

\begin{figure}[!ht]
    \centering
    \includegraphics[width=\linewidth]{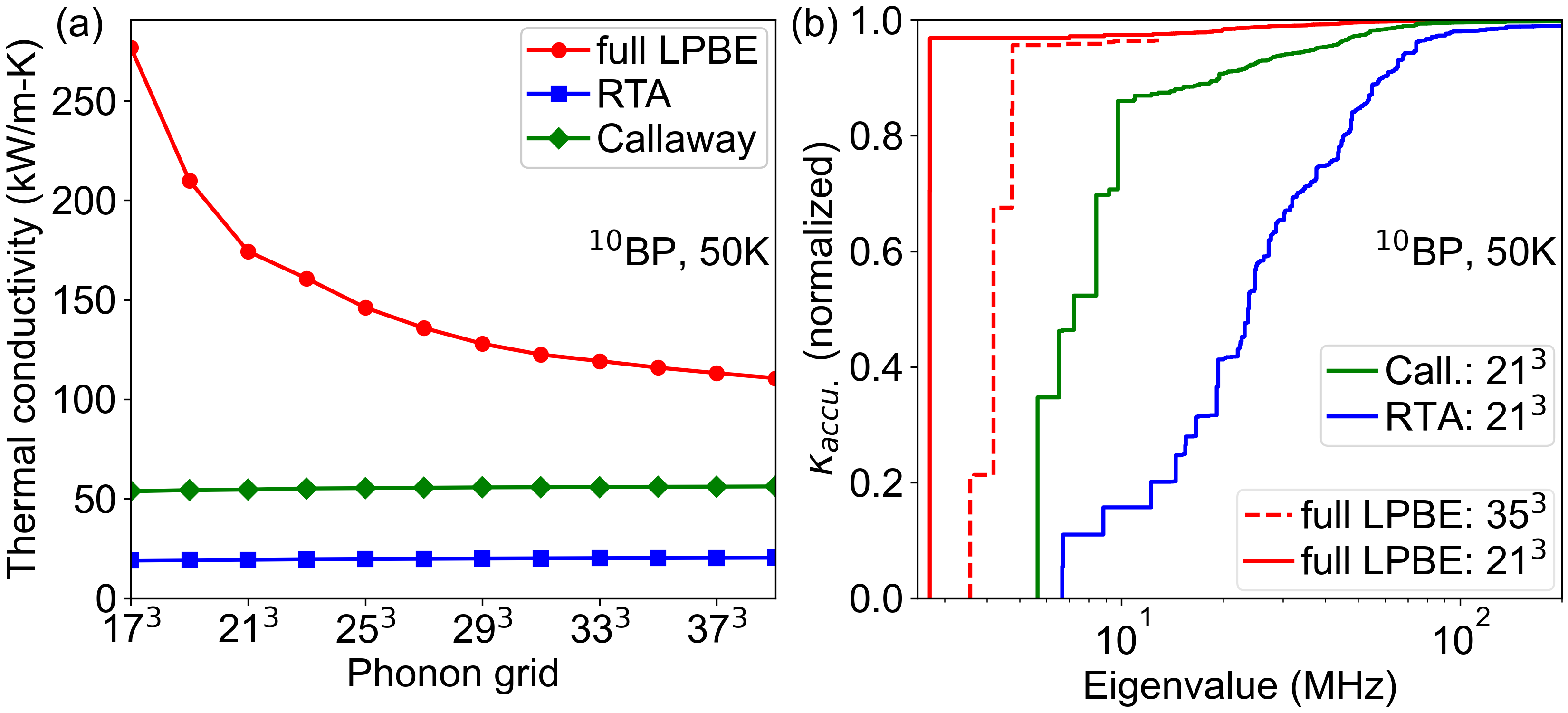}
    \caption{\textbf{(a)} Three phonon-limited $\kappa$ vs. Brillouin zone (BZ) discretization from the full LPBE solution, the Callaway approximation and the RTA for $^{10}$BP at 50~K. The RTA- and the Callaway-$\kappa$ converge for a coarse discretization ($17^3$) whereas the full LPBE-$\kappa$ requires a finer discretization ($>35^3$). \textbf{(b)} Normalized cumulative $\kappa$ of $^{10}$BP at 50~K vs. the eigenvalues of the collision matrix of the full LPBE, the Callaway approximation and the RTA at a $21^3$ BZ discretization. Also shown is the cumulative $\kappa$ vs. the eigenvalues of the full LPBE collision matrix at a finer BZ discretization ($35^3$), where $\kappa_{LPBE}$ converged to $\lesssim 17\%$ of the extrapolated value at infinite BZ discretization. Convergence of cumulative $\kappa$ from the Callaway approximation with BZ discretization is shown in Supplementary Fig. S2 (c). The LPBE solution shows strong BZ discretization-dependence of $\kappa$ in stark contrast to that from the RTA and the Callaway approximation.}
    \label{fig:B10P_50K_grid_dependent_plots}
\end{figure}

However, obtaining the eigenmodes of $\mb{\Omega}$ for ultrahigh-$\kappa$ material, where hydrodynamic transport is most likely to occur, presents a practical computational challenge. As shown in Fig.~\ref{fig:B10P_50K_grid_dependent_plots}~(a), the calculated $\kappa_{LPBE}$ of a prototypical ultrahigh-$\kappa$ material~\cite{zhu_vapor-flux_2026} --- isotopically pure boron phosphide ($^{10}$BP) --- converges slowly with BZ discretization. Since the cost of diagonalization is $\sim\mathcal{O}\left(N^3\right)$ for an $N\times N$ matrix, complete diagonalization of $\mb{\Omega}$ for a fine (e.g., 35$^3$) BZ discretization is prohibitively expensive. Although Krylov solvers such as ARPACK~\cite{lehoucq_arpack_1998} can compute few eigenmodes of $\mb{\Omega}$ with the smallest eigenvalues more efficiently than full diagonalization~\cite{malviya_efficient_2025}, the number of eigenmodes needed to obtain a converged $\kappa_{LPBE}$ is not known apriori and must be determined by trial-and-error, thus making the diagonalization approach unsuitable for rapid screening of materials for phonon hydrodynamics.

Interestingly, the predicted decreasing $\kappa_{LPBE}$ with BZ discretization in Fig.~\ref{fig:B10P_50K_grid_dependent_plots} (a) correlates with the decreasing spectral contributions of individual eigenmodes of $\mb{\Omega}$ to $\kappa_{LPBE}$ in Fig.~\ref{fig:B10P_50K_grid_dependent_plots} (b). From Fig.~\ref{fig:B10P_50K_grid_dependent_plots} (b), we find that exactly three nearly-degenerate eigenmodes contribute to $\kappa_{LPBE}$ entirely for a  21$^3$ BZ discretization, while the spectral $\kappa_{LPBE}$ is more distributed across three different sets of eigenmodes with different eigenvalues for a 35$^3$ discretization, reflecting a weaker hydrodynamic signature in the finer discretization. This conclusion is also strengthened by the smaller eigenvalues of the eigenmodes contributing to $\kappa_{LPBE}$ for the 21$^3$ calculation compared to the 35$^3$ result, thus deviating farther away from the ideal drifting characteristic of vanishing eigenvalues in the latter. Thus, with $\kappa_{RTA}$ being nearly independent of the BZ discretization [Fig.~\ref{fig:B10P_50K_grid_dependent_plots} (a)], the ratio $\kappa_{LPBE}/\kappa_{RTA}$, which is a measure of the relative strength of $\mb{\Omega}^{\left(N\right)}$ and $\mb{\Omega}^{\left(U\right)}$ as discussed earlier, correlates with the strength of hydrodynamic signatures in Fig.~\ref{fig:B10P_50K_grid_dependent_plots} (b) as the BZ discretization is refined. Therefore, this ratio, obtained using an iterative solution of the linear system in Eq.~\ref{eq:steady_state_lpbe} in a computationally inexpensive manner, is a reliable indicator for phonon hydrodynamics without the computational challenge associated with the diagonalization of $\mb{\Omega}$.

\begin{figure}[!ht]
    \centering
    \includegraphics[width=\linewidth]{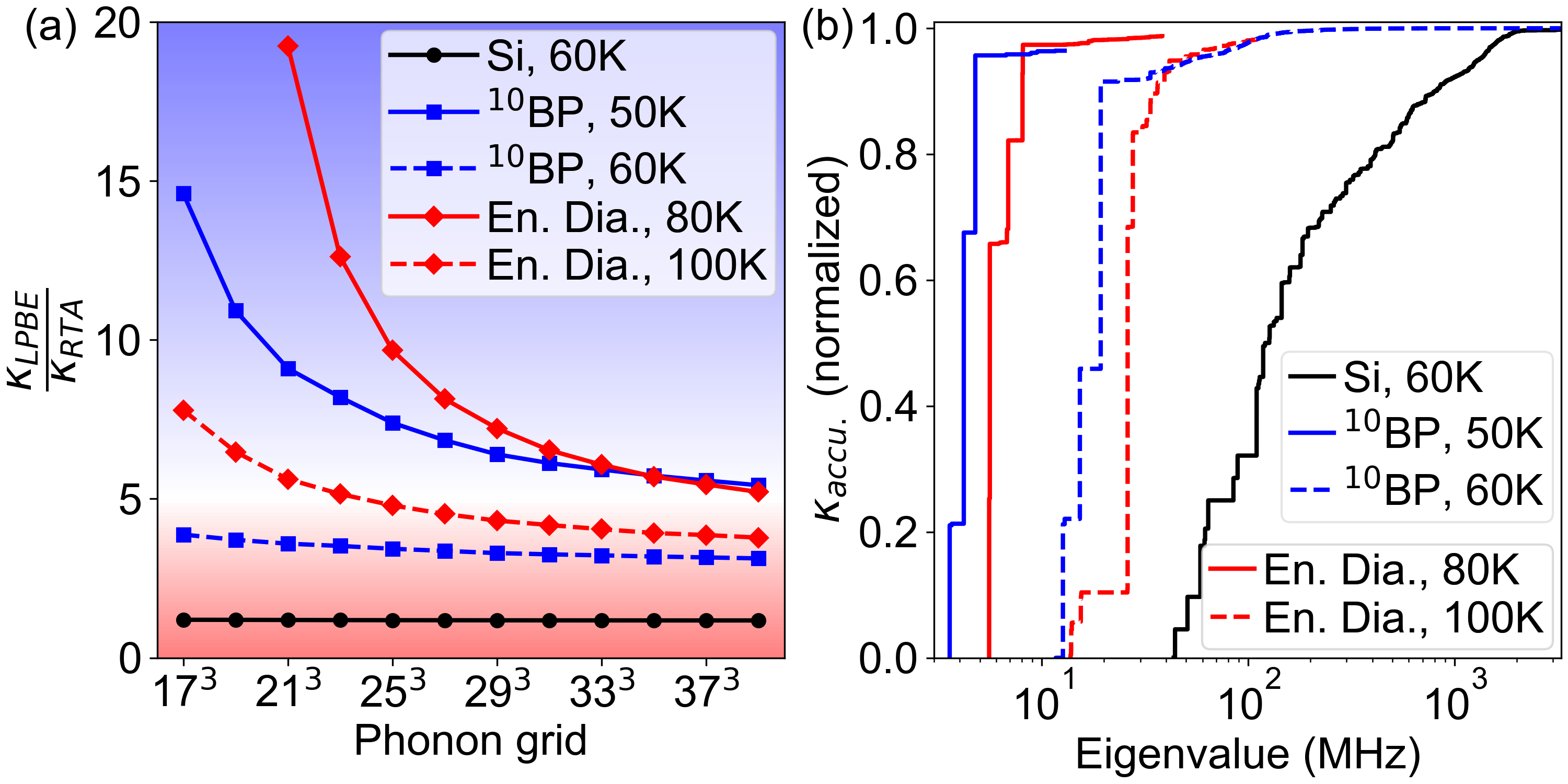}
    \caption{\textbf{(a)} $\kappa_{LPBE}/\kappa_{RTA}$ for Si at 60~K, $^{10}$BP at 50~K and 60~K and enriched diamond (En. Dia.) at 80~K and 100~K at different BZ discretizations. The converged values of $\kappa_{LPBE}/\kappa_{RTA}$ with respect to the BZ discretization correlates with the strength of hydrodynamic signatures and is illustrated by the background color transition from red to blue. \textbf{(b)} Cumulative $\kappa$ vs. eigenvalues for Si at 60~K for a $17^3$ BZ discretization, enriched diamond at 80 K and 100 K for a $35^3$ BZ discretization and $^{10}$BP at 50~K and 60 K at  $35^3$ and $21^3$ BZ discretizations respectively. While the cumulative $\kappa$ spectrum for Si even at a low temperature of 60 K is predominantly diffusive with small individual contributions from the eigenmodes, it shows significantly smaller diffuse contributions and large drifting contributions from individual eigenmodes resembling a strongly hydrodynamic signature~\cite{pitaevskii_physical_2012} in $^{10}$BP at 60 K and enriched diamond at 100 K, which are further amplified at lower temperatures.}
    \label{fig:new_second_sound_condition_3d_mtrl}
\end{figure}

To test our hypothesis, we plot $\kappa_{LPBE}/\kappa_{RTA}$ vs. BZ discretization for isotopically pure $^{10}$BP and isotopically enriched diamond at different temperatures in Fig.~\ref{fig:new_second_sound_condition_3d_mtrl} (a), and the spectral contribution to $\kappa_{LPBE}$ from the eigenmodes of $\mb{\Omega}$ calculated for the BZ discretization-converged $\kappa_{LPBE}$ for these materials in Fig.~\ref{fig:new_second_sound_condition_3d_mtrl} (b). We have also included the results for Si at 60 K as a reference material for which the ratio $\kappa_{LPBE}/\kappa_{RTA} \sim 1$ for all BZ discretizations in Fig.~\ref{fig:new_second_sound_condition_3d_mtrl} (a) whose $\kappa_{LPBE}$ accumulation has contributions from a broad spectrum of eigenmodes of $\mb{\Omega}$ with eigenvalues ranging from $\sim 40$ MHz to $\sim 1$ GHz in Fig.~\ref{fig:new_second_sound_condition_3d_mtrl} (b), thus representing a diffusive heat flow regime. For $^{10}$BP at 60 K and enriched diamond at 100 K, $\kappa_{LPBE}/\kappa_{RTA}$ converges to a value of $\sim 4$ for BZ discretizations of $21^3$ and $35^3$ respectively, and shows a stronger hydrodynamic signature with larger fractional contributions from individual eigenmodes with smaller eigenvalues compared to Si at 60 K in Fig.~\ref{fig:new_second_sound_condition_3d_mtrl} (b). More importantly, the fraction of the eigenmodes that individually contribute to less than 2\% of the overall $\kappa_{LPBE}$ --- which we refer to as diffusive eigenmodes --- is significantly smaller in $^{10}$BP (60 K) and enriched diamond (100 K) than in Si (60 K). These features of large drifting and small diffusive contributions of the eigenmodes to $\kappa_{LPBE}$ indicate a strongly hydrodynamic phonon transport in these materials. When the temperature is lowered further, for $^{10}$BP (50 K) and enriched diamond (80 K), the ratio $\kappa_{LPBE}/\kappa_{RTA}$ requires finer BZ discretizations to converge ($\gtrsim$35$^{3}$), due to reduced thermal occupation of large $\mb{q}$ phonons, which results in sharply weakened U-scattering~\cite{malviya_failure_2023}. At such low temperatures, these materials exhibit increasing $\kappa_{LPBE}/\kappa_{RTA}$, with the contributions to $\kappa_{LPBE}$ from the eigenspectrum of $\mb{\Omega}$ approaching the limiting case of pure drifting hydrodynamics when the U-processes are vanishingly small. Furthermore, the fractional contribution to $\kappa$ from the diffusive eigenmodes of $\mb{\Omega}$ is less than 10\% of the total value when the ratio $\kappa_{LPBE}/\kappa_{RTA}$ exceeds five in the materials considered here.

\begin{figure}[!ht]
    \centering
    \includegraphics[width=\linewidth]{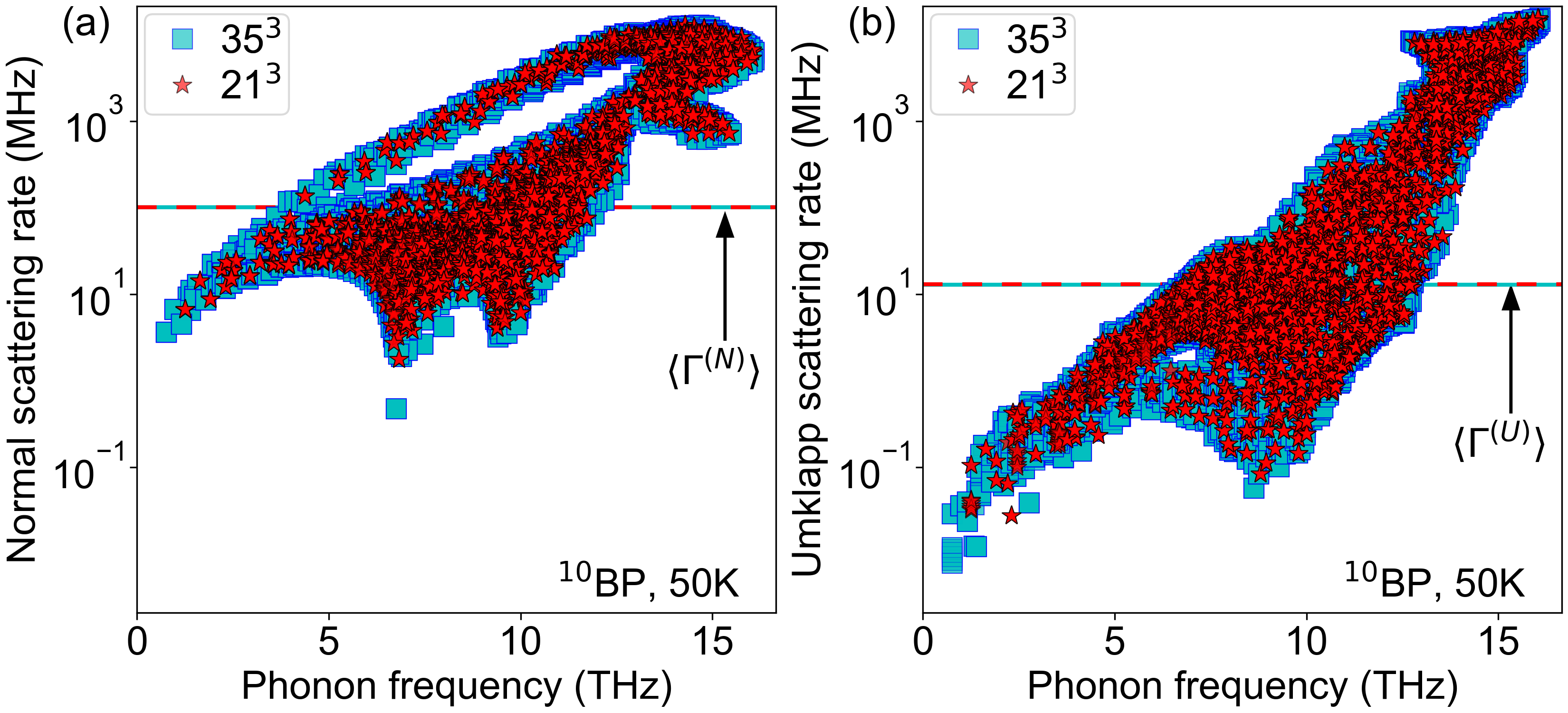}
    \caption{The three phonon scattering rates of (a) Normal and (b) Umklapp processes for $^{10}$BP at 50 K. These scattering rates and their thermally-averaged values ($\langle \Gamma^{N/U} \rangle$) show insignificant change with refinement of the Brillouin zone (BZ) discretization from $21^3$ and $35^3$, and so, do not correlate with the BZ discretization-dependence of the spectral contribution of $\kappa_{LPBE}$ in terms of the eigenmodes of the collision operator shown in Fig.~\ref{fig:B10P_50K_grid_dependent_plots} (b).}
    \label{fig:B10P_50K_grid_dependent_scattering_rates}
\end{figure}

We note that the weakening hydrodynamic signatures and decreasing indicator ratio with increasing BZ discretization [Fig.~\ref{fig:new_second_sound_condition_3d_mtrl}] cannot be explained by comparing the scattering rates for N- and U-processes alone. As shown in Fig.~\ref{fig:B10P_50K_grid_dependent_scattering_rates}, these scattering rates ($\Gamma_{\lambda}$) and their thermally-averaged values --- $\expval{\Gamma^{\left( N/U \right)}} = \frac{\sum_{\lambda} \Gamma_{\lambda}^{\left( N/U \right)} C_{\lambda}}{\sum_{\lambda} C_{\lambda}}$, indicated by the horizontal lines in Fig.~\ref{fig:B10P_50K_grid_dependent_scattering_rates}, are independent of the BZ discretization, and therefore, cannot account for the BZ discretization-dependence of $\kappa_{LPBE}/\kappa_{RTA}$. 

Furthermore, approximate solutions of the LPBE using the diagonal terms of $\mb{\Omega}$ (i.e., the N- and the U-scattering rates) only --- $\kappa_{RTA}$ and $\kappa_{Callaway}$ --- are much lower and nearly BZ discretization-independent, as shown for $^{10}$BP at 50~K in Fig.~\ref{fig:B10P_50K_grid_dependent_plots} (a). The latter feature originates from the BZ discretization-independence of N- and U-scattering rates [Fig.~\ref{fig:B10P_50K_grid_dependent_scattering_rates}]. To explain the former feature, following Ref.~\cite{malviya_failure_2023}, we define effective collision matrices $\Omega_{\lambda\lambda_1}^{\left(RTA\right)} = \frac{\Delta_{\lambda\lambda_1}}{\tau_\lambda}$ and $\Omega_{\lambda\lambda_1}^{\left(Callaway\right)} = \frac{\Delta_{\lambda\lambda_1}}{\tau_\lambda} - \mb{q}\cdot\tilde{\mb{q}}'\frac{\sqrt{f^0_\lambda\left(f^0_\lambda + 1\right)}\sqrt{f^0_{\lambda_1}\left(f^0_{\lambda_1} + 1\right)}}{\tau_\lambda^{\left(N\right)}\tau_{\lambda'}^{\left(N\right)}}$ for the RTA and the Callaway approximations to the LPBE respectively, with $\tilde{\mb{q}} = \left[\sum_{\lambda_1}\mb{q}_1\otimes\mb{q}_1\frac{f^0_{\lambda_1}\left(f^0_{\lambda_1} + 1\right)}{\tau_{\lambda_1}^{\left(N\right)}}\right]^{-1}\mb{q}$ and $\tau_\lambda^{\left(N\right)}$ being the relaxation time of the phonon mode $\lambda$ due to N-processes only. We see from Fig.~\ref{fig:B10P_50K_grid_dependent_plots} (b) that the eigenvalues of these effective collision matrices are larger, and show smaller individual fractional contributions to $\kappa$ relative to those of the LPBE collision matrix $\mb{\Omega}$, indicating that the eigenmodes of the RTA and the Callaway collision matrices are more diffusive in nature relative to those of $\mb{\Omega}$, leading to their lower $\kappa$. Therefore, these approximate models that rely solely on the diagonal elements of $\mb{\Omega}$ are unlikely to robustly predict the strength of phonon hydrodynamic signatures, even when N-scattering rates are much larger than U-scattering rates [Fig.~\ref{fig:B10P_50K_grid_dependent_scattering_rates} and Supplementary Fig.~S3]. Furthermore, these approximate solutions may not be unique, since the classification of N- and U-processes depends on the choice of the first BZ~\cite{lindsay_perspective_2019, duan_hydrodynamically_2022, markov_hydrodynamic_2018}. Since $\kappa_{LPBE}$ and $\kappa_{RTA}$ are obtained from the total $\mb{\Omega}$ and the total scattering rates respectively, they are independent of this choice, thereby improving the reliability of $\kappa_{LPBE}/\kappa_{RTA}$ towards predicting phonon hydrodynamics.

\begin{figure}[!ht]
    \centering
    \includegraphics[width=\linewidth]{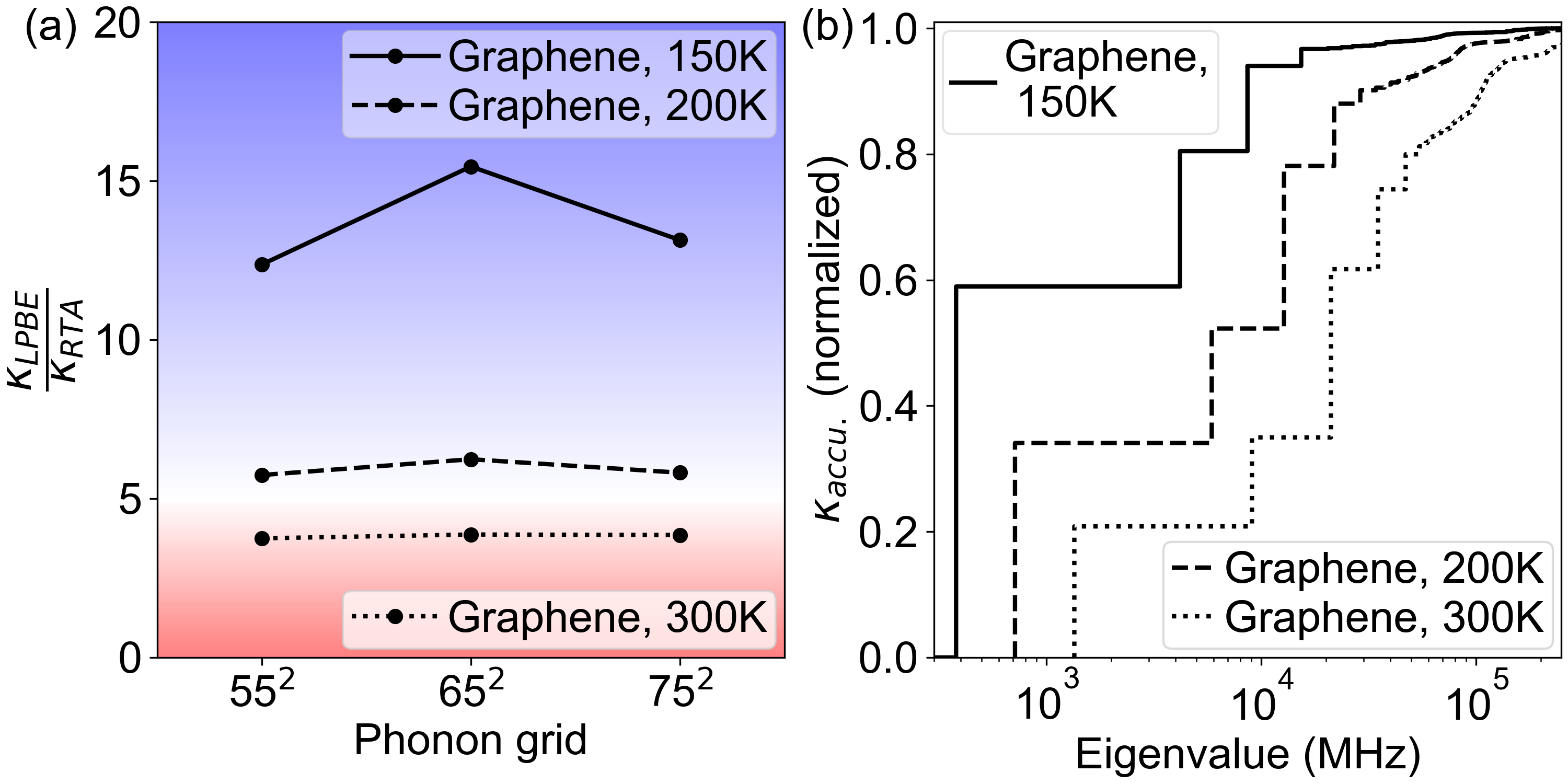}
    \caption{\textbf{(a)} $\kappa_{LPBE}/\kappa_{RTA}$ for suspended monolayer graphene at 150~K, 200~K and 300~K. The corresponding spectral accumulation functions for $\kappa_{LPBE}$ vs. eigenvalues of $\mb{\Omega}$, computed with three-phonon as well as four-phonon processes after including anharmonic renormalization for the flexural phonons, are shown in \textbf{(b)}. The ratio $\kappa_{LPBE}/\kappa_{RTA}$ increases with decreasing temperature in \textbf{(a)}, and correlates with strengthening hydrodynamic signatures in \textbf{(b)}.}
    \label{fig:new_second_sound_condition_2d_mtrl}
\end{figure}

We have recently shown in Ref.~\cite{ravichandran_elasticity_2026} that suspended monolayer graphene must exhibit strong hydrodynamic signatures even at temperatures as high as 150~K, driven by the anharmonic renormalization~\cite{ravichandran_low_2026} and higher-order four-phonon scattering of the flexural phonons --- the out-of-plane vibrational modes. Here we find from Fig.~\ref{fig:new_second_sound_condition_2d_mtrl} (a) that $\kappa_{LPBE}/\kappa_{RTA}$ of graphene is $\sim 5$ at 200~K and $\sim 13$ at 150~K, while $< 4$ at 300~K. The corresponding spectral contributions to $\kappa$ from the drifting eigenmodes in Fig.~\ref{fig:new_second_sound_condition_2d_mtrl} (b) evolve from $\sim 80\%$ at 300~K to $> 95\%$ at 150~K, while the diffusive contributions decrease from $\lesssim 20\%$ at 300~K to $< 5\%$ at 150~K, thus confirming that the ratio $\kappa_{LPBE}/\kappa_{RTA}$ is a reasonable indicator for the strength of phonon hydrodynamics in two-dimensional (2D) materials as well.

Finally, our results also shed light on the importance of careful convergence tests on the indicator ratio $\kappa_{LPBE}/\kappa_{RTA}$ while exploring materials and experimental conditions for hydrodynamic heat flow. We find that the eigenvalues of the drifting eigenmodes and their contribution to $\kappa_{LPBE}$ that distinguish the hydrodynamic regime from its diffusive counterpart, could be overpredicted [e.g., in enriched diamond at 100~K and $^{10}$BP at 50~K; see Supplementary Figs. S2 (a) and (b) respectively] at coarse discretizations of the BZ. However, these quantities achieve convergence at around the same BZ discretization [Supplementary Figs. S1 and S2] as the indicator ratio $\kappa_{LPBE}/\kappa_{RTA}$ in 2D as well as 3D crystals [Figs.~\ref{fig:new_second_sound_condition_2d_mtrl} and~\ref{fig:new_second_sound_condition_3d_mtrl} respectively]. Thus, the indicator ratio $\kappa_{LPBE}/\kappa_{RTA}$ is also a computationally inexpensive surrogate to determine the finest necessary discretization of the BZ to accurately predict hydrodynamic signatures, without the need for testing convergence of the computationally expensive eigenvalues of $\mb{\Omega}$ at finer BZ discretizations. Instead, the computationally expensive diagonalization of $\mb{\Omega}$ is required only for the final converged BZ discretization to obtain the spectral contributions to $\kappa_{LPBE}$. We also note that the indicator ratio $\kappa_{LPBE}/\kappa_{RTA}$ cannot distinguish between the specific types of the hydrodynamic heat flow regime, such as Poiseuille~\cite{simoncelli_generalization_2020, dragasevic_viscous_2026, di_lucente_vortices_2026, huang_observation_2023, machida_observation_2018}, Ziman~\cite{kawabata_phonon_2025}, and second-sound heat flow regimes~\cite{lee_hydrodynamic_2020}, since it does not directly encode the dominant momentum-dissipation mechanism or other characteristic signatures associated with these regimes, such as the parabolic heat-flux profile associated with Poiseuille flow~\cite{li_reexamination_2022} and an oscillatory temperature response from second sound~\cite{malviya_bridging_2026}. However, as we demonstrate in this manuscript, it can reliably help in screening for materials and experimental conditions conducive to support a collectively drifting phonon population that is central these different types of hydrodynamic regimes.

In summary, we uncover a direct correlation between the ratio $\kappa_{LPBE}/\kappa_{RTA}$ and the microscopic signatures of phonon hydrodynamics emerging from the spectral properties of the phonon collision operator, that establishes the former as a reliable low-cost indicator for hydrodynamic heat flow. We show that this connect emerges from the ability of the indicator ratio to capture the contributions of the off-diagonal part of the phonon collision operator to phonon hydrodynamics for ultrahigh-$\kappa$ materials. On the other hand, approaches that utilize only the diagonal part --- the phonon scattering rates, cannot reliably predict the strength of hydrodynamic transport. We also show that convergence studies with respect to the BZ discretization is essential to quantitatively capture the hydrodynamic signatures in a material arising from the drift eigenmodes of $\mb{\Omega}$, for which the indicator ratio $\kappa_{LPBE}/\kappa_{RTA}$ can serve as a surrogate at a fraction of the computational cost associated with the exact diagonalization of $\mb{\Omega}$. By justifying the use of the ratio $\kappa_{LPBE}/\kappa_{RTA}$ as a low-cost indicator for phonon hydrodynamics, our work, thus, opens up pathways to accelerate data-driven computational searches of ideal materials and experimental conditions for realizing such unconventional non-diffusive heat flow regimes that transcend the conventional Fourier's law of heat diffusion.

See supplementary materials for convergence of the spectral-$\kappa$ vs. eigenvalues of $\mb{\Omega}$ with BZ discretization for different materials and a summary of the first-principles methodology used here.

This work was supported by the Core Research Grant (CRG) No. CRG/2022/009160, and the Mathematical Research Impact Centric Support (MATRICS) Grant No. MTR/2022/001043 from the Science and Engineering Research Board, India, by the Advanced Research Grant (ARG) No. ANRF/ARG/2025/007160/ENS from the Anusandhan National Research Foundation, India, and by the Indo-Japan Cooperative Science Programme (IJCSP) grant TPN No. 139097 from the Department of Science and Technology (India) and the Japanese Society for the Promotion of Science (Japan). N.K.R. acknowledges the Infosys Foundation for the Young Investigator Award. N.M. gratefully acknowledges the Prime Minister's Research Fellowship (PMRF) grant no. PMRF-02-01036.

\section*{Author declarations}
\subsection*{Conflict of Interest}
The authors have no conflicts to disclose.

\subsection*{Author Contributions}
N.K.R. originated the research idea. N.M. developed the computational framework and performed the calculations. N.M. and N.K.R. developed the theory, analyzed the results, and wrote the manuscript.

\subsection*{Data Availability}
All formulations and computational optimizations necessary to perform the calculations are described in Refs.~\cite{ravichandran_unified_2018, malviya_failure_2023, malviya_efficient_2025, ravichandran_low_2026}.

%
\end{document}